\documentclass{jpp}

\usepackage{graphicx}
\usepackage{natbib}

\usepackage{bm}
\usepackage{epsfig}

\usepackage{amsmath}

\usepackage{amsfonts}
\usepackage{amssymb}

\newcommand\etal{\mbox{\textit{et al.}}}

\newsavebox{\astrutbox}
\sbox{\astrutbox}{\rule[-5pt]{0pt}{20pt}}

\newcommand\de{\partial}
\newcommand\bv{\mathbf{v}}

\title[Solar wind collisional heating]{Solar wind collisional heating}

\author[O. Pezzi]{Oreste Pezzi\thanks{Email address for correspondence: oreste.pezzi@fis.unical.it}}

\affiliation{Dipartimento di Fisica, Universit\`a della Calabria, 87036 Rende (CS), Italy.}

\pubyear{}
\volume{}
\pagerange{}
\date{?; revised ?; accepted ?. - To be entered by editorial office}

\begin{document}
\maketitle

\begin{abstract}
To properly describe heating in weakly collisional turbulent plasmas such as the solar wind, inter-particle collisions should be taken 
into account. Collisions can convert ordered energy into heat by means of irreversible relaxation towards the thermal equilibrium. 
Recently, Pezzi et al. ({\em Phys. Rev. Lett.}, vol.  116, 2016, p. 145001) showed that the plasma collisionality is enhanced by the 
presence of fine structures in velocity space. Here, the analysis is extended by directly comparing the effects of the fully nonlinear 
Landau operator and a linearized Landau operator. By focusing on the relaxation towards the equilibrium of an out of equilibrium 
distribution function in a homogeneous force-free plasma, here it is pointed out that it is significant to retain nonlinearities in the 
collisional operator to quantify the importance of collisional effects. Although the presence of several characteristic times associated 
with the dissipation of different phase space structures is recovered in both the cases of the nonlinear and the linearized operators, the 
influence of these times is different in the two cases. In the linearized operator case, the recovered characteristic times are 
systematically larger than in the fully nonlinear operator case, this suggesting that fine velocity structures are dissipated slower if 
nonlinearities are neglected in the collisional operator.

\end{abstract}

\begin{PACS}
Authors should not enter PACS codes directly on the manuscript, as these must be chosen during the online submission process and 
will then be added during the typesetting process (see http://www.aip.org/pacs/ for the full list of PACS codes)
\end{PACS}

\section{Introduction}
\label{sec:intro}

Since the beginning of the last Century, many theoretical efforts have been performed to model natural and laboratory plasmas. One of the 
first attempts to describe the interplanetary medium and its interaction with the planetary magnetospheres was conducted by S. Chapman and 
V.C.A. Ferraro \citep{chapman30,chapman31}, widely considered the fathers of the Magnetohydrodynamics (MHD) theory. Their main intuition was 
to treat plasmas, approximated by neutral conducting fluids, as self-consistent media. One of the basic assumptions of this framework is 
that inter-particle collisions are sufficiently strong to maintain a local thermodynamical equilibrium, e.g. the particle velocity 
distribution function (VDF) is close to the equilibrium Maxwellian shape. This approach is still widely adopted to analyze plasma dynamics 
at large scales and many models have been developed to study the features of the MHD turbulence \citep{elsasser50, chandrasekhar56, 
iroshnikov64, kraichnan65, moffatt78, parker79, dobrowolny80a, dobrowolny80b, NgBhattacharjee, MatthaeusEA99, verdini09, bruno13, howes13, 
pezzi16b}. One of the most studied natural plasmas is the solar wind, which is the high temperature, low density, supersonic flow 
emitted from the solar atmosphere. The solar wind is a strongly turbulent flow: the typical Reynolds number is about $Re\approx 10^{5}$ 
\citep{matthaeus05}; fluctuations are broadband and often exhibit a power-law spectra; several indicators of intermittency are 
routinely observed \citep{bruno13,matthaeus15}. Despite the solar wind is usually approached in terms of MHD turbulence, spacecraft {\it 
in-situ} measurements reveal much complex features, which go beyond the fluid MHD approach. Once the energy is transferred by turbulence 
towards smaller scales close to the ion inertial scales, kinetic physics signatures are often observed \citep{SahraouiEA07, GaryEA10, 
AlexandrovaEA08, bruno13}. The particle VDF often displays a distorted out-of-equilibrium shape characterized by the presence of 
non-Maxwellian features such as temperature anisotropies, particle beams along the local magnetic field direction, rings-like structures  
\citep{marsch06, kasper08, maruca11, maruca13, he15}. The principal models to take into account kinetic effects are based on the assumption 
that the plasma is {\it collisionless}, e.g. collisions are far too weak to produce any significant effect on the plasma dynamics 
\citep{daughton09, ParasharPP09, camporeale11, valentini11, servidio12, greco12, perrone12, valentini14, franci15, servidio15, 
valentini16}. 

We would point out that, in order to comprehend the heating mechanisms of the solar wind, collisional effects should be considered. Indeed 
collisions are the unique mechanism able to produce irreversible heating from a thermodynamic point of view. Furthermore, to show that 
collisions can be neglected, the shape of the particle VDF is usually assumed to be close to the equilibrium Maxwellian 
\citep{spitzer56, hernandez85, maruca13}. This approximation may result problematic for weakly-collisional turbulent plasmas, where kinetic 
physics strongly distorts the particle VDFs and produces fine structure in velocity space. Collisional effects, which explicitly depend on 
gradients in velocity space, may be enhanced by the presence of these small scale structures in velocity space \citep{pezzi16a} (here after 
Paper I). Indeed, in Paper I we showed that the the collisional thermalization of fine velocity structures occurs on much smaller 
times with respect to the usual Spitzer-Harm time \citep{spitzer56} $\nu_{SH}^{-1}$ (being $\nu_{SH}\simeq 8 \times (0.714 \pi n e^4 
\ln \Lambda)/(m^{0.5} (3 k_B T)^{3/2})$, where $n$, $e$, $\ln \Lambda$, $m$, $k_B$ and $T$ are respectively the particle number density, the 
unit electric charge, the Coulombian logarithm, the Boltzmann constant and the plasma temperature). The smallest characteristic times may 
be comparable with the characteristic times of other physical processes. Therefore, collisions could play a significant role into the 
dissipation of strong gradients in the VDF, thus contributing to the plasma heating.

In this paper we focus on the importance of retaining nonlinearities in the collisional operators. In particular, by means of numerical 
simulations of a homogeneous force-free plasma, we describe the collisional relaxation towards the equilibrium of an initial VDF which 
exhibits strong non-Maxwellian signatures. Collisions among particles of the same species are here modeled through the fully nonlinear 
Landau operator and a linearized Landau operator. A detailed comparison concerning the effects of the two operator indicates that retaining 
nonlinearities in the collisional integral is crucial to give the proper importance to collisional effects. Indeed, both operators are able 
to highlight the presence of several characteristic times associated with the dissipation of fine velocity structures. However, the 
magnitude of these times is different if nonlinearities are neglected: in the linearized operator case, the characteristic times are 
systematically larger compared to the case of the fully nonlinear operator. This indicates that, when nonlinearities are not taken into 
account in the mathematical form of the collisional operator, fine velocity structures are dissipated much slower. Results here described 
support the idea that to properly quantify the enhancement of collisional effects and, hence, to correctly compare collisional times with 
other dynamical times, it is important to adopt nonlinear collisional operators. 

We would remark that, since the Landau operator is demanding from a computational perspective, self-consistent high-resolution 
simulations cannot be currently afforded and we are forced to restrict to the case of a force-free homogeneous plasma, where both force and 
advection terms have been neglected.  This approximation represents a caveat of the work here presented and future studies will be devoted 
to the generalization of the results here shown to the self-consistent case. 

The paper is organized as follows: in Sec. \ref{sec:SWheat} the solar wind heating problem is revisited in order to address and motivate 
our work. Then, in Sec. \ref{sec:numres} we give a brief description of the numerical codes and the adopted methods of analysis. Numerical 
results of our simulations are also reported and discussed in detail. Finally, in Sec. \ref{sec:concl} we conclude and summarize.

\section{Solar wind heating: a huge problem}
\label{sec:SWheat}

As introduced above, the solar wind is a weakly collisional, strongly turbulent medium \citep{bruno13}. Several observations indicate that 
the solar wind is incessantly heated during its travel through the heliosphere: the temperature decay along the radial distance is indeed 
much slower than the decay expected within adiabatic models of the wind expansion \citep{marsch82,goldstein96,marino08,cranmer09}. 
Therefore, some local heating mechanisms must play a significant role to supply the energy needed to heat the plasma. Numerous scenarios 
have been proposed to understand the plasma heating and a long-standing debate about which processes are preferred is still waiting for a 
clear and definitive answer [See \citet{bruno13} and references therein]. Among these processes, it is widely known that the turbulence 
efficiently contributes to the local heating of solar wind \citep{sorriso07, marino08, sahraoui09}, since it can efficiently transfer a 
significant amount of energy towards smaller scales, where dissipative mechanisms are at work. In fact, in a turbulent flow, much more 
energy is transferred towards smaller scales with respect to a laminar flow: the ratio between the energy transfer flux due to turbulence at 
a certain scale with respect to the heating production due to dissipation at the same scale is proportional to the Reynolds number $Re$, 
thus indicating that the energy transfer towards smaller scales gets more efficient as the flow becomes more turbulent. 

In the simple neutral fluid scenario, the cascade is arrested once that the dissipative scale is reached \citep{frisch95}. On the other 
hand, the cascade evolves in a more complex way in a plasma: the presence of other processes (for example dispersion and kinetic effects) 
strongly modify the cascade before reaching the dissipative scale. A relatively wide agreement has been achieved about the importance of 
turbulence for transferring energy towards smaller scales. Instead, many scenarios have been proposed to explain the transition from the 
inertial range towards the kinetic scales and the nature of dissipative processes. These scenarios are often based on the ``collisionless'' 
assumption, that is justified by the fact that the Spitzer-Harm collisional time \citep{spitzer56} is much larger than other dynamical 
times. We would remark that two important caveats should be considered. 

First, any mechanism which does not consider collisions is not able to describe the last part of the heating process, namely the heat 
production due to the irreversible dissipation of phase space structures and the approach towards the thermal equilibrium. For example, 
several mechanisms (e.g. nonlinear waves) can indeed increase the particle temperature, evaluated as the second order moment of the particle 
distribution function, by producing non-Maxwellian features as beams of trapped particles. However, this temperature growth due to 
the beam production does not represent a temperature growth in the thermodynamic sense, because the beam presence makes the system out of 
equilibrium. The particles beam can be instead interpreted as a form of free energy stored into the VDF. This energy is not in general 
converted into heat by means of irreversible processes but it can be also transformed in other forms of ordered energy (e.g. through 
micro-instabilities) \citep{lesur14}. Collisions are the unique mechanism able to degrade this information into heat by approaching the 
thermal equilibrium, thus producing heating in the general thermodynamic and irreversible sense. 

Second, the evaluation of the Spitzer-Harm collisional time strictly assumes that the VDF shape is close to the equilibrium Maxwellian. 
This assumption may not be held in the solar wind \citep{marsch06,servidio15}, where VDFs shape is strongly perturbed by kinetic 
turbulence. In this direction, by focusing on the collisional relaxation in a homogeneous force-free plasma where collisions are modeled 
with the fully nonlinear Landau operator \citep{landau36}, we recently showed that fine velocity structures are dissipated much faster than 
global non-thermal features such as temperature anisotropy (Paper I). The entropy production due to the relaxation of the VDF towards the 
equilibrium occurs on several characteristic times. These characteristic times are associated with the dissipation of particular velocity 
space structures and can be much smaller than the Spitzer-Harm time \citep{spitzer56}, this indicating that collisions could effectively 
compete with other processes (e.g. micro-instabilities). In this perspective, high-resolution measurements of the particle VDF in the solar 
wind are crucial for a proper description of the heating problem \citep{vaivads16}.

In principle the combination of the turbulent nature of the solar wind with its weakly collisionality may constitute a new scenario to 
describe the solar wind heating. In fact, turbulence is able to transfer energy towards smaller scales. Then, when kinetic scales are 
reached, since the plasma is weakly collisional, the VDF becomes strongly distorted and exhibits non-Maxwellian features, such as beams, 
anisotropies, ring-like structures \citep{belmont08, chust09, servidio12, servidio15}. The presence of strong gradients in velocity 
space tends to naturally enhance the effect of collisions, which - ultimately - may become efficient for dissipating these structures and 
for producing heat. 

Based on these last considerations, numerous studies have been recently conducted in order to take into account collisional effects in a 
weakly collisional plasma such as the solar wind \citep{filbet02, bobylev13, pezzi13, pezzi14a, escande15, pezzi15b, pezzi16a, 
banonnavarro16, hirvijoki16, tigik16}, where collisions are usually introduced by means of a collisional operator at the right hand-side of 
the Vlasov equation. The choice of the {\it proper} collisional operator remains an open problem. Several derivations from first 
principles (e.g. Liouville equation) indicate that the most general collisional operators for plasmas are the Lenard-Balescu operator 
\citep{lenard60, balescu60} or the Landau operator \citep{landau36,akhiezer86}. Both operators are nonlinear ``Fokker-Planck''-like 
operators which involve velocity space derivatives and three-dimensional integrals. The Landau operator introduces an upper cut-off of the 
integrals at the Debye length to avoid the divergence for large impact parameters, while the Balescu-Lenard operator solves this divergence 
in a more consistent way through the dispersion equation. Therefore, the Balescu-Lenard operator is more general compared to the Landau 
operator from this point of view. However, both operators are derived by assuming that the plasma is not extremely far from the thermal 
equilibrium. Hence, both operators could lack the description of inter-particle collisions in a strongly turbulent system. The numerical 
approach of operators is also much more difficult for the Balescu-Lenard operator with respect to the Landau operator, because it involves 
the evaluation of dispersion function. Finally, we would also point out that, as far as we know, an explicit derivation of the Boltzmann 
operator for plasmas starting by the Liouville equation does not exist \citep{villani02}. Despite the adoption of the Boltzmann operator 
for describing collisional effects in plasmas is questionable from a theoretical perspective, it still represents a valid options since 
Boltzmann and Fokker-Planck like operators such as the Landau one are intrinsically similar \citep{landau36,bobylev13}.

The computational cost to evaluate both the Landau and Balescu-Lenard operators numerically is huge: for $N$ gridpoints along each 
direction of the $3D$--$3V$ numerical phase space ($3D$ in physical space and $3D$ in velocity space), the computation for the Landau 
operator would require about $N^9$ operations at each time step.  In fact, for each point of the six-dimensional grid, a 
three-dimensional integral must be computed. To avoid this numerical complexity, several simplified operators have been proposed. We may 
distinguish these simpler operators in two classes. The first type of operators, as the Bathanar-Gross-Krook \citep{bgk54,livi86} and the 
Dougherty operators \citep{dougherty64,dougherty67,pezzi15a}, models collisions in the realistic three-dimensional velocity space by 
adopting a simpler structure of the operator. The second class of collisional operators works instead in a reduced, one-dimensional velocity 
space assuming that the dynamics mainly occur in one direction. Although this approach is quite ``unphysical'' (collisions naturally act in 
three dimensions), these operators can satisfactorily model collisions in laboratory plasmas devices, such as the Penning-Malmberg traps, 
where the plasma is confined into a long and thin column and the dynamics occurs mainly along a single direction \citep{anderson07a, 
anderson07b, pezzi13}.

\section{Numerical approach and simulation results}
\label{sec:numres}

As described above, to highlight the importance of nonlinearities present in the collisional operator, here we compare the effects of the 
fully Landau operator with a model of linearized Landau operator,  obtained by simplifying the structures of the Landau operator 
coefficients. We restrict to the case of a force-free homogeneous plasma and we just model collisions between particles of the same 
species. Our interest is in fact to understand how collisional effects change when the mathematical kernel of the collisional operator is 
modified. Based on these assumptions, we numerically integrate the following dimensionless collisional evolution equations for the particle 
distribution function $f(\bv,t)$: 
\begin{eqnarray}
\frac{\de f(\bv,t)}{\de t} & = & \pi \left(\frac{3}{2}\right)^{\frac{3}{2}}\frac{\de }{\de v_{i}}  \int d^3v' \ U_{ij}
(\mathbf{u}) \left[ f(\bv',t)\frac{\de f(\bv,t)}{\de v_{j}} - f(\bv,t) \frac{\de f(\bv',t)}{\de v'_{j}} \right] \ ,   
\label{eq:lanNL} \\ 
\frac{\de f(\bv,t)}{\de t} & = & \pi \left(\frac{3}{2}\right)^{\frac{3}{2}}\frac{\de }{\de v_{i}}  \int d^3v' \ U_{ij}
(\mathbf{u}) \left[ f_0(\bv')\frac{\de f(\bv,t)}{\de v_{j}} - f(\bv,t) \frac{\de f_0(\bv')}{\de v'_{j}} \right] \ .  
\label{eq:lanLIN}
\end{eqnarray}
being $f$ normalized such that $\int d^3v f(\bv) =n=1$ and $U_{ij}(\mathbf{u})$
\begin{equation}
 U_{ij}(\mathbf{u})  = \frac{\delta_{ij}u^2 - u_i u_j}{u^3} \ ,
 \label{lanproj}
\end{equation}

where $\mathbf{u}=\bv -\bv'$, $u=|\mathbf{u}|$ and the Einstein notation is introduced. In Eqs. (\ref{eq:lanNL}--\ref{eq:lanLIN}), 
and from now on, time is scaled to the inverse Spitzer-Harm frequency $\nu_{SH}^{-1}$ \citep{spitzer56} and velocity to the particle thermal 
speed $v_{th}$. Details about the numerical solution of Eqs. (\ref{eq:lanNL}--\ref{eq:lanLIN}) can be found in Refs. \citep{pezzi15a, 
pezzi16a}. In Eq. (\ref{eq:lanLIN}), $f_0(\bv)$ is the three-dimensional Maxwellian distribution function associated with the initial 
condition of our simulations $f(\bv,t=0)$ and built in such a way that density, bulk velocity and temperature of the two distributions 
$f(\bv,t=0)$ and $f_0(\bv)$ are equal. The two equations clearly differ because Eq. (\ref{eq:lanLIN}) is a linearized model of Eq. 
(\ref{eq:lanNL}). The operator described in Eq. (\ref{eq:lanLIN}) has been in fact obtained by linearizing the coefficients of the 
Landau operator. Although this linear operator does not represent the exact linearization of the Landau operator, the procedure here 
adopted for linearizing the operator (i.e. simplifying only the Fokker-Planck coefficients) is commonly adopted. In the following, we will 
note that the simulations performed with the linearized operator thermalize to the same final VDF and produce also the same total entropy 
growth of the nonlinear simulations. This suggests that the term which is not included in the form collisional operator (whose 
Fokker-Planck coefficients  depend on $(f-f_0)(\bv')$) is not extremely relevant in the global thermalization of the system. This 
approximation corresponds to retain the gradients related to the out-of-equilibrium structures but to neglect their contribute to the 
integral in the $\bv'$ space. In other words, here we locally consider gradients but we neglect their contribute to the global Fokker-Planck 
coefficients.

When simulations are completed, we perform the following multi-exponential fit \citep{curtis70,pezzi16a} of the entropy growth $\Delta S$ to 
point out the presence of several characteristic times:
\begin{equation}
 \Delta S (t) = \sum_{i=1}^{K} \Delta S_i \left( 1 - e^{-t/\tau_i} \right) \ ,
 \label{eqfit}
 \end{equation}
$\tau_i$ being the $i$--th characteristic time, $\Delta S_i$ the growth of entropy related to the characteristic time $\tau_i$ and $K$ is 
evaluated through a recursive procedure. This procedure has been already adopted in Paper I to highlight the importance of fine velocity 
structures in the entropy growth. In the following subsections, we report and describe the results of the simulations performed with 
two different initial distribution functions, already adopted in Paper I. The first initial condition concerns the presence of 
non-Maxwellian signatures due to a strongly nonlinear wave - an Electron Acoustic Wave (EAW) \citep{holloway91, kabantsev06, valentini06, 
anderegg09a, anderegg09b, johnston09, valentini12} - in the core of the distribution function. The EA waves here excited are quite different 
from another type of electron acoustic fluctuations which occur in a plasma composed by two components at different temperature 
\citep{watanabe77} and can be also observed in the Earth's magnetosphere \citep{tokar84,lu05}. The EAWs here excited are undamped waves 
whose phase speed is close to the thermal speed. It has been shown that, in the usual theory of the equilibrium Maxwellian plasma, these 
waves are strongly damped; while they can survive if the distribution function is locally modified (and exhibits a flat region) around the 
wave phase velocity. To generate the nonlinearity in the distribution function and let these waves survive, external drivers are usually 
adopted to force the plasma. EAWs are also characterized by the presence of phase space Bernstein-Green-Kruskal (BGK) structures 
\citep{bgk57} in the core of the electron distribution function, associated with trapped particle populations. The second initial 
distribution is instead a typical VDF recovered in hybrid Vlasov-Maxwell simulations of solar wind decaying turbulence \citep{servidio12, 
valentini14, servidio15}. The two simulations from which we selected our initial VDF are quite different. Indeed, in the first case, the 
out-of-equilibrium structures present in the initial VDF are due to the wave-particle interaction with the EAW, which is an almost 
monochromatic (few excited wavenumbers), electrostatic wave. On the other hand, in the second case, the initial distribution function has 
been strongly distorted due to the presence of an electromagnetic, turbulent cascade.

\subsection{First case study: wave-particle interactions and collisions} 

\begin{figure}
\epsfxsize=\textwidth \centerline{\epsffile{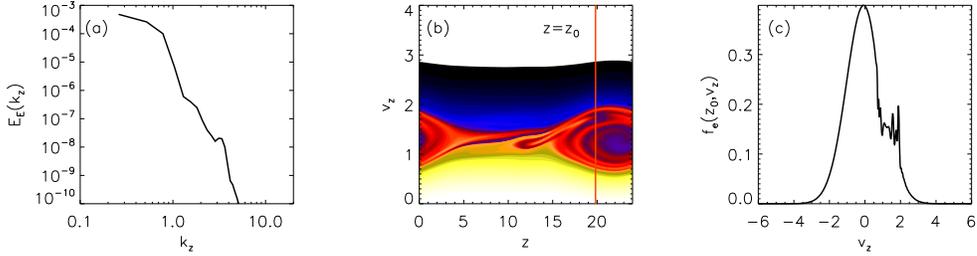}}   
\caption{(Color online) (a) Power spectral density of the electric energy $E_E(k_z)$ as a function of the wavenumber $k_z$. (b) Contour 
plot of $f_e(z,v_z)$ at the time instant when the EAWs is fully developed. The red line represents the coordinate $z=z_0$ where the cut is 
performed. (c) Profile of $f_e(z_0,v_z)$ as a function of $v_z$. }
\label{fig:fd0}
\end{figure}

The first initial condition here adopted, which is a three-dimensional VDF that evolves according to Eqs. (\ref{eq:lanNL}--\ref{eq:lanLIN}) 
in the three-dimensional velocity space, has been designed as follows. We separately performed a $1D$--$1V$ Vlasov-Poisson simulation of a 
electrostatic plasma composed by kinetic electrons and motionless protons whose resolution, in the $z-v_z$ phase space domain, is $N_z=256$, 
$N_{v_z}=1601$. In order to excite a large amplitude EAW, we forced the system with an external sinusoidal electric field, which has been 
adiabatically turned on and off to properly trigger the wave.  Fig. \ref{fig:fd0}(a) reports the power spectral density of the electric 
energy $E_E(k_z)$ as a function of the wavenumber $k_z$, evaluated at the final time instant of the Vlasov-Poisson simulation (where the EAW 
is fully developed). Few wavenumbers are significantly excited and the EAW is almost monochromatic. The features of the electric 
fluctuations spectrum are reflected into the shape of the distribution function, which is locally distorted around the phase speed and 
present a clear BGK hole, as reported in Fig. \ref{fig:fd0}(b). Since the gridsize in velocity space is quite small in the current 
simulation, relatively small velocity scales are dynamically generated during the simulation by wave-particle interaction.

\begin{figure}
\epsfxsize=10cm \centerline{\epsffile{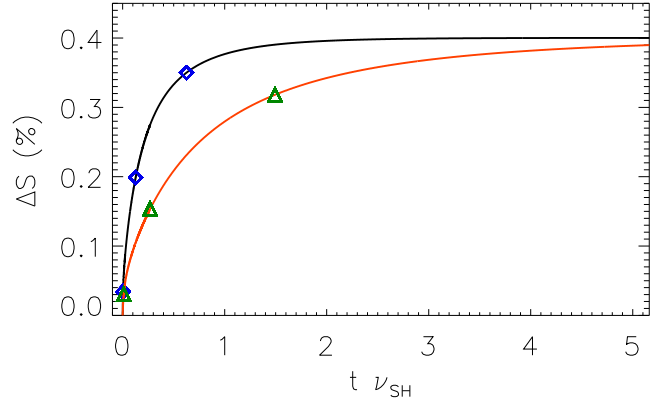}}   
\caption{(Color online) Time history of $\Delta S$ in the case of the fully nonlinear Landau operator (black) and the linearized 
Landau operator (red). Blue diamonds indicate the time instants $t=T_{nl,1}=\tau^{nl}_1$, $t=T_{nl,2}=\tau^{nl}_1+\tau^{nl}_2$ and
$t=T_{nl,3}=\tau^{nl}_1+\tau^{nl}_2+\tau^{nl}_3$; the green triangles refer to $t=T_{lin,1}=\tau^{lin}_1$, 
$t=T_{lin,2}=\tau^{lin}_1+\tau^{lin}_2$ and $t=T_{lin,3}=\tau^{lin}_1+\tau^{lin}_2+\tau^{lin}_3$. }
\label{fig:entr}
\end{figure}

Then, we selected the spatial point $z_0$ in the numerical domain [red vertical line in Fig. \ref{fig:fd0}(b)], where this BGK-like phase 
space structure displays its maximum velocity width, and we considered the velocity profile $\hat{f}_e(v_z)=f_e(z_0,v_z)$, whose shape as a 
function of $v_z$ is reported in Fig. \ref{fig:fd0}(c). $\hat{f}_e$ is highly distorted due to nonlinear wave-particle interactions and 
exhibits sharp velocity gradients (bumps, holes, spikes around the resonant speed). Finally, by evaluating the density $n_e$, the bulk speed 
$V_e$ and the temperature $T_e$ of $\hat{f}_e$, we built up the three-dimensional VDF $f(v_x,v_y,v_z)=f_{M}(v_x)f_{M}(v_y)\hat{f}_e(v_z)$, 
which represents our initial condition, being $f_M$ the one-dimensional Maxwellian associated with $\hat{f}_e$. We remark that this VDF does 
not exhibit any temperature anisotropy but it still exhibits strong non-Maxwellian deformations along $v_z$, due to the presence of trapped 
particles, which make the system far from thermal equilibrium. The three-dimensional velocity domain is here discretized by 
$N_{v_x}=N_{v_y}=51$ and $N_{v_z}=1601$ gridpoints in the region $v_i=[-v_{max},v_{max}]$, being $v_{max}=6v_{th}$ and $i=x,y,z$, while 
boundary conditions assume that the distribution function is set equal to zero for $|v_j|>v_{max}$.

Since no temperature anisotropies are present, the evolutions of the total temperature and of the temperatures along each direction are 
trivial, the total temperature is preserved. On the other hand, the evolution of the entropy variation $\Delta S=S(t)-S(0)$ ($S=-\int 
f\ln{f} d^3v$) gives information about the approach towards equilibrium. The time history of $\Delta S$ obtained with the nonlinear Landau 
operator (black) and with the linearized Landau operator (red) is showed in Fig. \ref{fig:entr}. Since the initial condition and the 
equilibrium Maxwellian reached under the effect of collisions is the same for both operators, the total entropy growth $\Delta S$ is the 
same in the two cases. In other words, the free energy contained in the out-of-equilibrium structures in the initial VDF produces the same 
entropy growth in absolute terms but the growth occurs on different time scales in the two cases. Indeed, in the nonlinear operator case the 
entropy grows much faster ($1\div2 \nu_{SH}^{-1}$) compared to the linearized operator case ($4\div 5 \nu_{SH}^{-1}$).

\begin{figure}
\epsfxsize=\textwidth \centerline{\epsffile{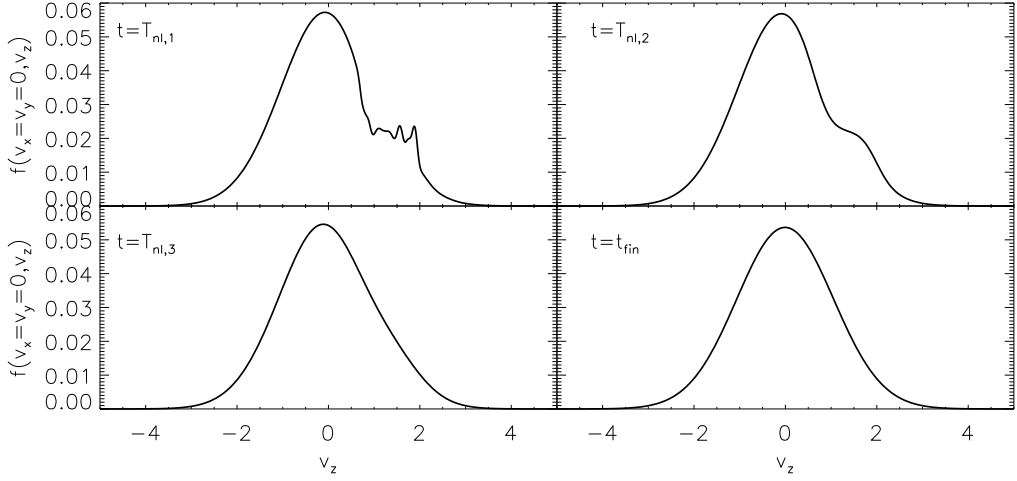}}   
\caption{Distribution function $f(v_x=0,v_y=0,v_z)$ as a function of $v_z$, obtained in the case of the fully nonlinear Landau 
operator. Panels from (a) to (d) respectively display the time instants $t=T_{nl,1}=\tau^{nl}_1$ (a), $t=T_{nl,2}=\tau^{nl}_1+\tau^{nl}_2$ 
(b), $t=T_{nl,3}=\tau^{nl}_1+\tau^{nl}_2+\tau^{nl}_3$ (c) and $t=t_{fin}$ (d). Results for the linearized Landau operator are qualitatively 
the same as here with the primary difference being that $\tau^{nl}_j$ is substantially smaller than $\tau^{lin}_j$ for each value of $j$.}
\label{fig:fdNL}
\end{figure}

To quantify the presence of several characteristic times, we perform the multi exponential fit of Eq. (\ref{eqfit}) on the entropy growth 
curves reported in Fig. \ref{fig:entr}. The analysis of the growth recovered in the fully nonlinear Landau operator indicates that three 
different characteristic times are recovered in the entropy growth:
\begin{itemize}
 \item $\tau^{nl}_1 = 3.5 \cdot 10^{-3}\, \nu_{SH}^{-1} \rightarrow \Delta S^{nl}_1 / \Delta S_{tot} = 13 \% $
 \item $\tau^{nl}_2 = 1.3 \cdot 10^{-1}\, \nu_{SH}^{-1} \rightarrow \Delta S^{nl}_2 / \Delta S_{tot} = 42 \% $
 \item $\tau^{nl}_3 = 4.9 \cdot 10^{-1}\, \nu_{SH}^{-1} \rightarrow \Delta S^{nl}_3 / \Delta S_{tot} = 40 \% $
\end{itemize}
As discussed in Paper I, the presence of several characteristic times is associated with the dissipation of different velocity space 
structures. Fig. \ref{fig:fdNL} reports $f(v_x=v_y=0,v_z)$ as a function of $v_z$ at the time instants $t=T_{nl,1}=\tau^{nl}_1$ (a), 
$t=T_{nl,2}=\tau^{nl}_1+\tau^{nl}_2$ (b), $t=T_{nl,3}=\tau^{nl}_1+\tau^{nl}_2+\tau^{nl}_3$ (c) and $t=t_{fin}$ (d), These time instants are 
displayed in Fig. \ref{fig:entr} with blue diamonds. After the time $t=T_{nl,1}=\tau^{nl}_1$ (a), steep spikes visible in Fig. 
\ref{fig:fd0}(b) are almost completely smoothed out; then, at time $t=T_{nl,2}=\tau^{nl}_1+\tau^{nl}_2$ (b), the remaining plateau region 
is significantly rounded off, only a gentle shoulder being left; finally, after a time $t=T_{nl,3}=\tau^{nl}_1+\tau^{nl}_2+\tau^{nl}_3$ (c), 
the collisional relaxation to equilibrium is completed for the most part. A small percentage $\simeq 5\%$ of the total entropy growth is 
finally recovered for larger times and corresponds to the final approach to the equilibrium Maxwellian (d).

By performing the same analysis for the linearized Landau operator case, three characteristic times are also recovered:
\begin{itemize}
 \item $\tau^{lin}_1 = 1.1 \cdot 10^{-2}\, \nu_{SH}^{-1} \rightarrow \Delta S^{lin}_1 / \Delta S_{tot} = 11 \% $
 \item $\tau^{lin}_2 = 2.7 \cdot 10^{-1}\, \nu_{SH}^{-1} \rightarrow \Delta S^{lin}_2 / \Delta S_{tot} = 23 \% $
 \item $\tau^{lin}_3 = 1.5 \;\;\;\;\;\;\;\;\;\;\; \nu_{SH}^{-1} \rightarrow \Delta S^{lin}_3 / \Delta S_{tot} = 63 \% $
\end{itemize}
These characteristic times are systematically larger than the times recovered in the nonlinear operator case. The shape of the 
distribution function after each characteristic time (not shown here) is quite similar to the shape recovered in the case of the fully 
nonlinear operator evolution. The process of dissipation of fine velocity structure is, hence, qualitatively similar if one adopts 
nonlinear or linearized operators. However, significant quantitative differences occur: similar profiles in velocity space are indeed 
reached at very different times, being the characteristic times recovered in the linearized case significantly larger (about $4\div5$ times) 
than the times recovered in the nonlinear operator case.

Therefore, from a qualitative point of view, both operators are able to recover the fact that fine velocity space structures are dissipated 
faster as their scale gets finer (i.e. as the velocity space gradients become stronger). However, fine velocity structures are dissipated 
slower by linearizing the collisional operator. Moreover, it is also worth mentioning that the amount of entropy growth associated with each 
characteristic time slightly changes by ignoring nonlinearities. For example, in the case of the fully nonlinear Landau operator, about 
$55\%$ of the total entropy growth is produced when the initial spikes and the successive flat plateau are dissipated. On the other hand, 
in the linearized operator case, only about the $30\%$ of the total entropy growth is associated with these processes.

\begin{figure*} 
 \centering
 \begin{minipage}{0.45 \textwidth}
  \epsfxsize=8cm \centerline{\epsffile{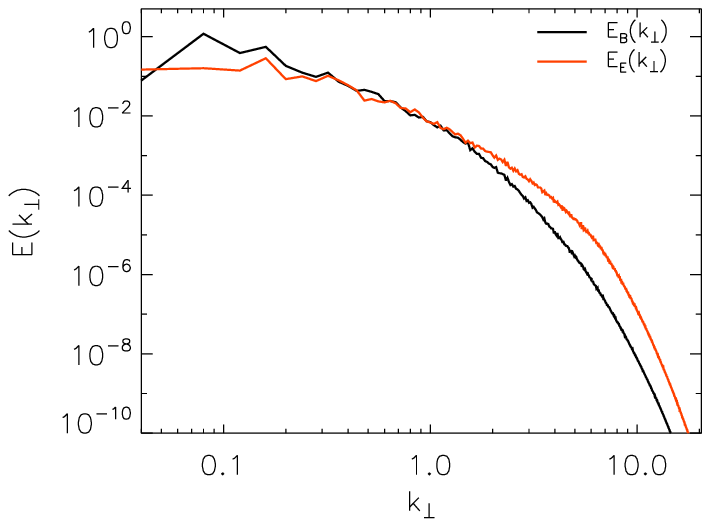}}   
 \end{minipage}
 \hfill
 \begin{minipage}{0.45 \textwidth}
  \epsfxsize=5.5cm \centerline{\epsffile{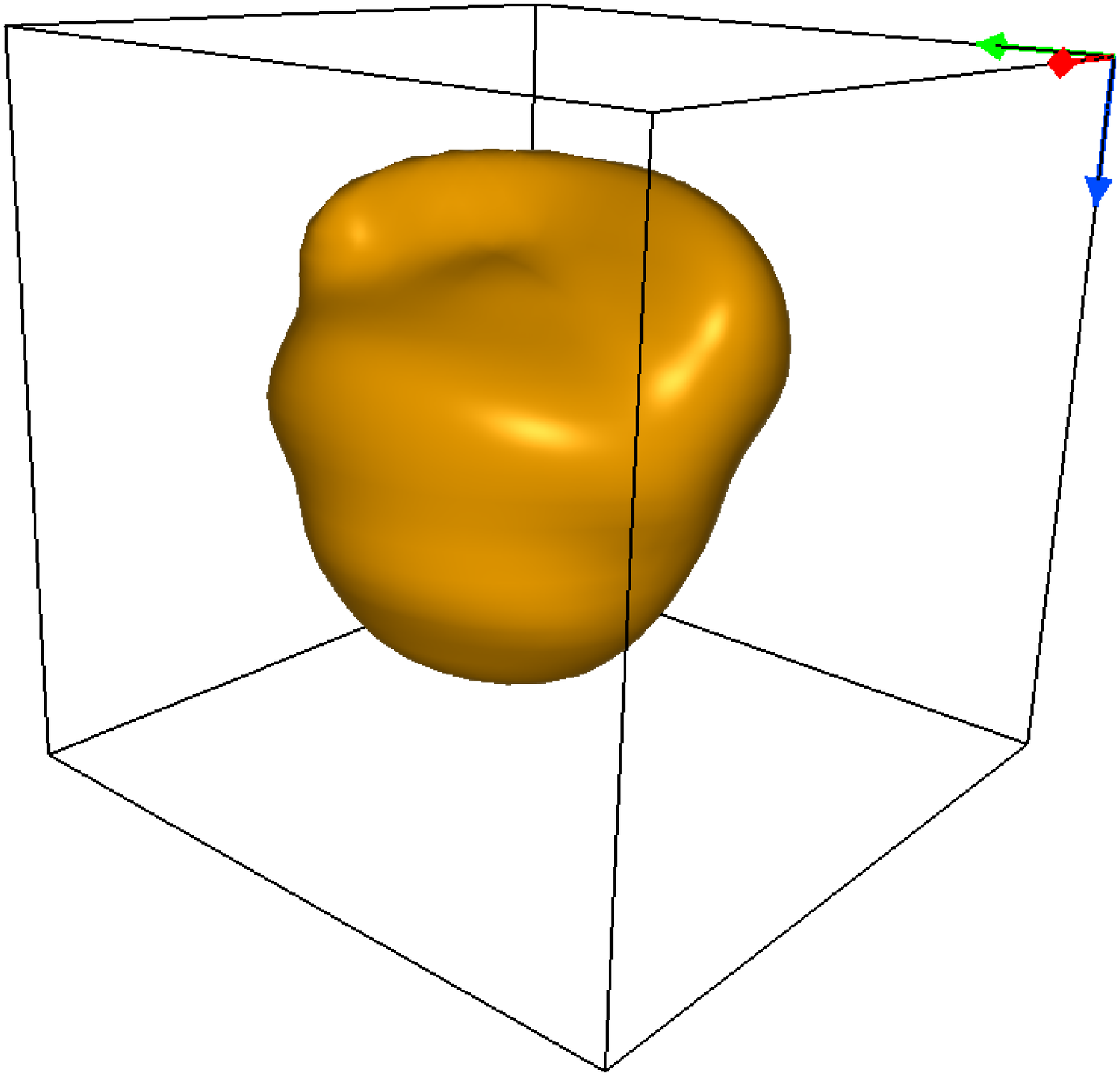}}   
 \end{minipage}
 \caption{(Color online) Left: Omni-directional power spectral densities of the magnetic energy $E_B(k_\perp)$ (black line) and of the 
electric energy $E_E(k_\perp)$ (red line) as a function of the perpendicular wavenumber $k_\perp$. PSDs have been evaluated at the time 
instant where the turbulent activity is maximum. Right: Iso-surface of the initial distribution function. Red, green and blue axes refer 
to $v_x$, $v_y$ and $v_z$, respectively.}
 \label{fig:TURBfd0}
\end{figure*}
 
\subsection{Second case study: kinetic turbulence and collisions} 

To support the scenario described in the previous section, here we focus on a second initial condition. This initial VDF has been selected 
from a $2D$--$3V$ hybrid Vlasov-Maxwell numerical simulation of decaying turbulence in solar wind like conditions 
\citep{valentini07,valentini14}. The hybrid Vlasov-Maxwell simulation, whose resolution is $N_x=N_y=512$ and 
$N_{v_x}=N_{v_y}=N_{v_z}=51$, is initialized with a out of the plane background magnetic field. Then, magnetic and bulk speed perturbations 
at large, MHD scales are introduced. As a result of nonlinear couplings among the fluctuations, the energy cascades towards smaller kinetic 
scales. Hence, the particle VDF strongly departs from the thermal equilibrium due to the presence of kinetic turbulence and exhibits a 
potato-like shape similar to the solar wind {\it in-situ} observations \citep{marsch06}. The omni-directional power spectral densities 
of the magnetic (black) and electric energy (line), evaluated at the time instant where the turbulent activity is maximum, are reported in 
Fig. \ref{fig:TURBfd0}(a): clearly a broadband spectrum is recovered. The iso-contour of the initial VDF, selected where non-Maxwellian 
effects are strongest \citep{servidio15}, is shown in Fig. \ref{fig:TURBfd0}(b). The VDF exhibits a hole-like structure in the upper part of 
the box and a thin ring-like structure in the bottom part of the box, while the VDF is clearly elongated on the $v_z$ direction. Compared to 
the first case study, the current distribution function reflects the presence of a spectrum of excited wavenumbers and it contains several 
kinds of distortions, not only concentrated around the resonant speed as in the previous case.

In Paper I we showed that, once velocity space gradients are artificially smoothed out through a fitting procedure, the presence of several 
characteristic times associated with the dissipation of fine velocity structures is definitively lost. Here, we instead compare the 
evolution towards the equilibrium of this initial condition under the effect of the fully nonlinear Landau operator [Eq. (\ref{eq:lanNL})] 
and the linearized Landau operator [Eq. (\ref{eq:lanLIN})]. The velocity domain is here discretized with $N_{v_x}=N_{v_y}=N_{v_z}=51$ 
points. Note that, compared to the first case study, the resolution is here smaller and it cannot be incremented due to the 
computational cost of the hybrid Vlasov-Maxwell code. Therefore, the quite small velocity scales recovered in the first case study (spikes 
around the resonant speed etc. etc.) are not present in this case.

\begin{figure}
\epsfxsize=10cm \centerline{\epsffile{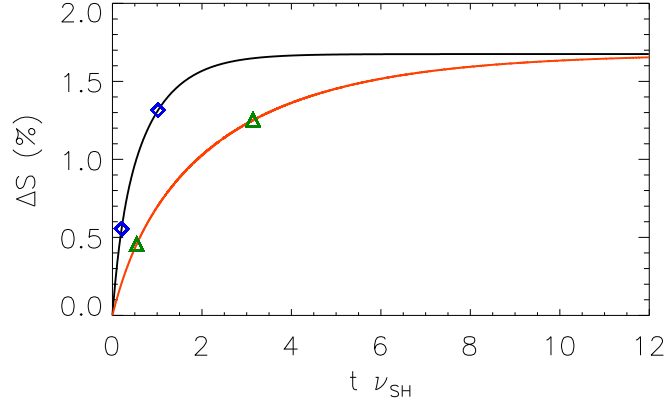}}   
\caption{(Color online) Time history of $\Delta S$ in the case of the fully nonlinear Landau operator (black) and the linearized 
Landau operator (red). Blue diamonds indicate the time instants $t=T_{nl,1}=\tau^{nl}_1$ and $t=T_{nl,2}=\tau^{nl}_1+\tau^{nl}_2$; the 
green triangles refer to $t=T_{lin,1}=\tau^{lin}_1$ and $t=T_{lin,2}=\tau^{lin}_1+\tau^{lin}_2$. }
\label{fig:TURBentr}
\end{figure}

Figure \ref{fig:TURBentr} reports the entropy growth obtained with the fully nonlinear Landau operator (black) and with its linearized 
version (red). The entropy growth is, also here, slower in the linearized operator case compared to the fully nonlinear operator case. To 
quantify the different evolution observed in Fig. \ref{fig:TURBentr}, we perform the multi-exponential fit \citep{curtis70,pezzi16a} 
described in Eq. (\ref{eqfit}).

\begin{figure*} 
 \centering
 \begin{minipage}{0.45 \textwidth}
  \epsfxsize=6cm \centerline{\epsffile{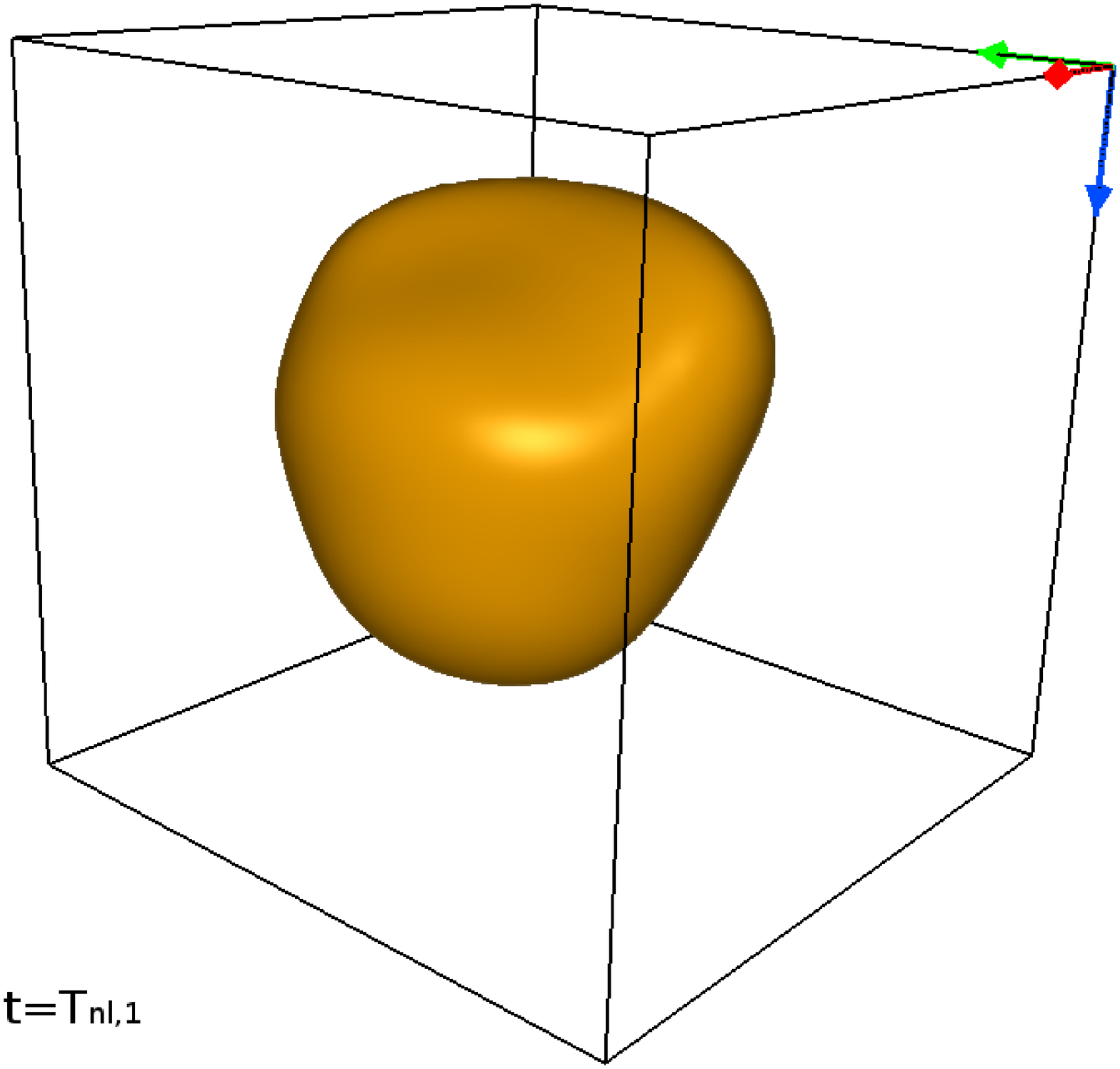}}   
 \end{minipage}
 \hfill
 \begin{minipage}{0.45 \textwidth}
  \epsfxsize=6cm \centerline{\epsffile{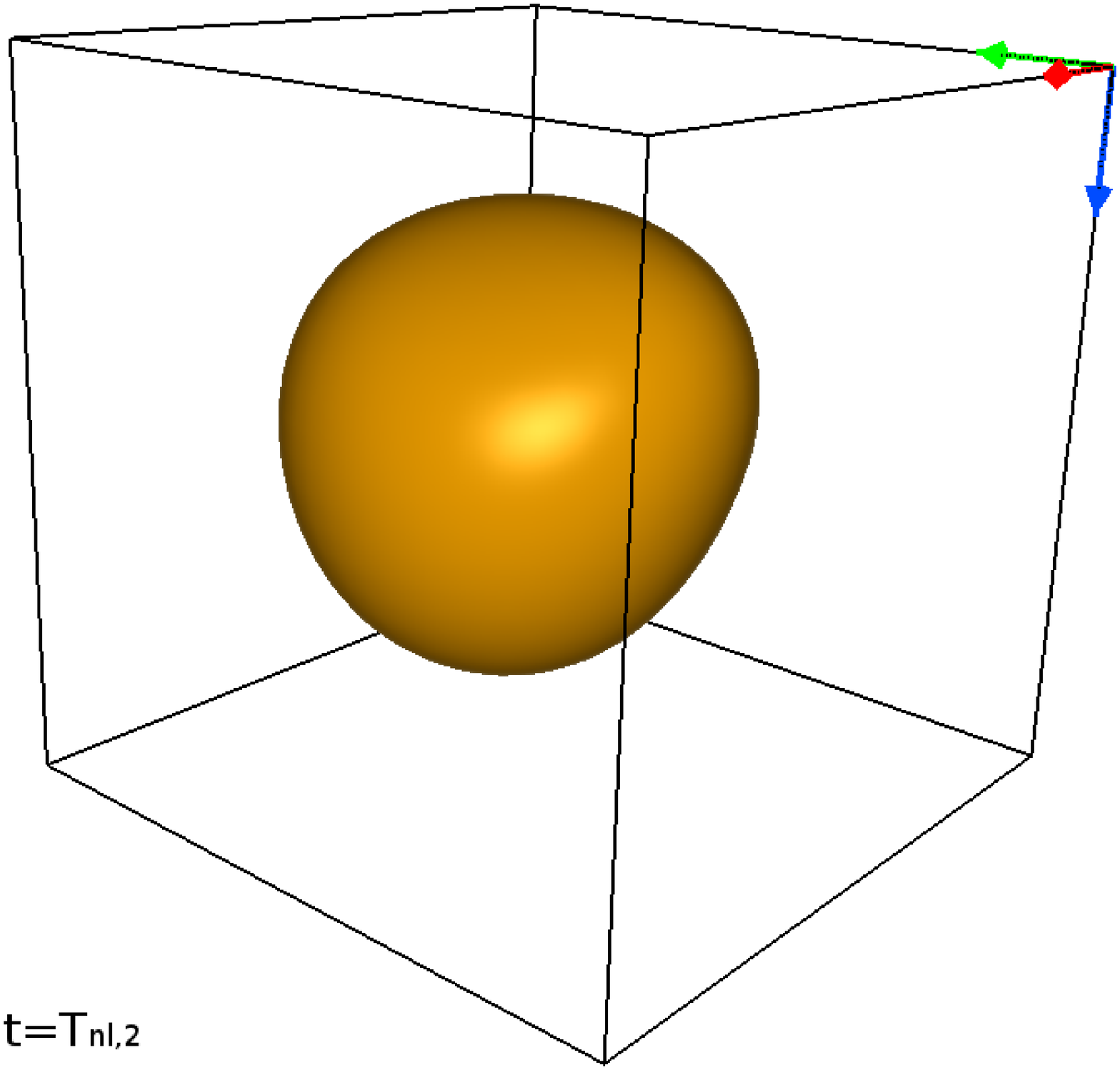}}   
 \end{minipage}
 \caption{(Color online) Iso-surface of the distribution function, obtained in the case of the fully nonlinear Landau operator. Left and 
right panels respectively display the time instants $t=T_{nl,1}=\tau^{nl}_1$ and $t=T_{nl,2}=\tau^{nl}_1+\tau^{nl}_2$. Red, green and blue 
axes refer to $v_x$, $v_y$ and $v_z$, respectively. The results for the linearized Landau operator are qualitatively the same as here 
with the primary difference being that $\tau^{nl}_j$ is substantially smaller than $\tau^{lin}_j$ for each value of $j$.}
 \label{fig:TURBVDFNL}
\end{figure*}

The analysis performed in the case of the fully nonlinear operator indicates that the entropy grows with two characteristic times:
\begin{itemize}
 \item $\tau^{nl}_1 = 0.20\, \nu_{SH}^{-1} \rightarrow \Delta S^{nl}_1 / \Delta S_{tot} = 26 \% $
 \item $\tau^{nl}_2 = 0.82 \, \nu_{SH}^{-1} \rightarrow \Delta S^{nl}_2 / \Delta S_{tot} = 74 \% $
\end{itemize}
As in the case described in the previous section, each characteristic time is associated with the dissipation of a different 
out-of-equilibrium features. Figure \ref{fig:TURBVDFNL} reports the iso-surface of the particle VDF at the time $t=T_{nl,1}=\tau^{nl}_1$ 
(left) and at the time $t=T_{nl,2}=\tau^{nl}_1+ \tau^{nl}_2$. At $t=T_{nl,1}$, the initial hole-like structure and the slight ring-like 
signature has been significantly smoothed out. Then, at $t=T_{nl,2}$, the approach towards the equilibrium is almost complete, being the VDF 
shape almost Maxwellian. Only a slight temperature anisotropy, which is finally thermalized in the late stage of the simulation, is still 
recovered. The approach towards the equilibrium confirms that small scale gradients are dissipated quite faster, while the final approach 
towards the equilibrium - concerning also the thermalization of temperature anisotropy - occurs on larger characteristic times. 

In the linearized operator case, two characteristic times are also recovered:
\begin{itemize}
 \item $\tau^{lin}_1 = 0.54 \, \nu_{SH}^{-1} \rightarrow \Delta S^{nl}_1 / \Delta S_{tot} = 16 \% $
 \item $\tau^{lin}_2 = 2.60 \, \nu_{SH}^{-1} \rightarrow \Delta S^{nl}_2 / \Delta S_{tot} = 84 \% $
\end{itemize}
As described for the first case study, these recovered characteristic times are systematically larger (about three times) compared to the 
times recovered in the fully nonlinear operator case. The amount of entropy growth associated with each characteristic time is also 
different, a smaller amount of entropy growth is indeed associated with the fastest characteristic time when nonlinearities are 
neglected. The results here described confirm the insights described in the previous section. The evolution obtained with the two operators 
is qualitatively similar: in both cases, several characteristic times are recovered in the entropy growth and these characteristic times are 
associated with the dissipation of different velocity space structures. However, the observed evolutions are different from a quantitative 
point of view: the recovered characteristic times are much different in the two cases, being significantly larger in the case of the 
linearized operator. 

As described in Paper I, in the first case study, much smaller characteristic times are in general recovered compared to the second 
case study, probably since the numerical resolution in the second case study is about $30$ times smaller compared to the first case study 
and the sharp velocity gradients present in the first case study [Fig. \ref{fig:fd0}(c)] are not accessible in the second case study [Fig. 
\ref{fig:TURBfd0}(b)]. The presence of finer velocity structures in the first case compared to the second case introduces smaller 
characteristic times.

\section{Conclusion}
\label{sec:concl}

To summarize, here we discussed in detail the importance of considering collisions in the description of the weakly collisional plasmas. 
Collisions are enhanced by the presence of fine velocity space structures, such as the ones naturally generated by kinetic turbulence 
in the solar wind; therefore, they could play a role into the conversion of VDFs free energy into heat, by means of irreversible processes. 

In particular, we focused on the importance of retaining nonlinearities in the collisional operator by performing a comparative analysis of 
the collisional relaxation of a out-of-equilibrium initial VDF. Collisions have been modeled by means of two collisional operators: the 
fully nonlinear Landau operator and a linearized Landau operator. Due to the demanding computational cost of the collisional integral, we 
restricted to the collisional relaxation in a force-free homogeneous plasma. Our results must be clearly extended to the more general, 
self-consistent case; however, performing a high-resolution collisional simulation cannot be currently afforded. 

The cases of study here analyzed indicate that both nonlinear and linearized collisional operators are able to detect the presence of 
several time scales associated with the collisional dissipation of small velocity scales in the particle VDF. A possible explanation of 
this behavior is that also the linearized operator involves gradients in its structure while it does not describe the ``second-order'' 
gradients related to the Fokker-Planck coefficients of the operator; therefore it is able to recover the presence of several characteristic 
times. The general message given in Paper I, namely the presence of sharp velocity space gradients speeds up the entropy growth of 
the system, is confirmed also in the case of the linearized operator: indeed, the fastest recovered characteristic times are significantly 
smaller than the common Spitzer-Harm collisional time \citep{spitzer56}. 

However, we would point out that the importance of the fine velocity structures is weakened if nonlinearities are ignored in the collisional 
operator. In the case of a linearized collisional operator, slower characteristic times are systematically recovered with respect to the 
nonlinear operator case. This indicates that, when one neglects the nonlinearities of the collisional integral, fine velocity structures are 
dissipated slower. Therefore, to properly address the role of collisions and to attribute them the correct relevance with respect to other 
physical processes \citep{matthaeus14,gary93,tigik16}, nonlinearities should be explicitly considered.

\section*{Acknowledgements}
Dr. O. Pezzi would sincerely thank Prof. P. Veltri, Dr. F. Valentini and Dr. D. Perrone for the fruitful discussions which significantly 
contributed to the construction of this work. Dr. O. Pezzi would also thank the anonymous Referees for their suggestions which improved the 
quality of this work. Numerical simulations here discussed have been run on the Fermi parallel machine at Cineca (Italy), within the 
Iscra--C project IsC26--COLTURBO and on the Newton parallel machine at University of Calabria (Rende, Italy). This work has been supported 
by the Agenzia Spaziale Italiana underthe Contract No. ASI-INAF 2015-039-R.O ``Missione M4 di ESA: Partecipazione Italiana alla fase di 
assessment della missione THOR''.

\bibliographystyle{jpp}

\end{document}